\documentclass[11pt]{article}

\newcommand{\BABARPubYear}    {00}

\newcommand{\BABARProcNumber} {24}
\newcommand{\SLACPubNumber} {8738}

\RequirePackage{xspace}

\def\Lbabar{\mbox{{\LARGE\sl B}\hspace{-0.15em}{\Large\sl A}\hspace{-0.07em}{\LARGE\sl B}\hspace{-0.15em}{\Large\sl A\hspace{-0.02em}R}}}
\def\lbabar{\mbox{{\large\sl B}\hspace{-0.4em} {\normalsize\sl A}\hspace{-0.03em}{\large\sl B}\hspace{-0.4em} {\normalsize\sl A\hspace{-0.02em}R}}}
\usepackage{relsize}
\def\babar{\mbox{\slshape B\kern-0.1em{\smaller A}\kern-0.1em
    B\kern-0.1em{\smaller A\kern-0.2em R}}}

\def\epem       {\ensuremath{e^+e^-}\xspace}

\def\s  {\ensuremath{s}\xspace}

\def\ccbar {\ensuremath{c\overline c}\xspace}
\def\b  {\ensuremath{b}\xspace}

\def\Kbar  {\kern 0.2em\overline{\kern -0.2em K}{}\xspace}

\def\KS    {\ensuremath{K^0_{\scriptscriptstyle S}}\xspace}

\def\Kzb   {\ensuremath{\Kbar^0}\xspace}
\def\KzKzb {\ensuremath{K^0 \kern -0.16em \Kzb}\xspace}

\def\Dbar  {\kern 0.2em\overline{\kern -0.2em D}{}\xspace}

\def\Dzb   {\ensuremath{\Dbar^0}\xspace}
\def\DzDzb {\ensuremath{D^0 {\kern -0.16em \Dzb}}\xspace}

\def\Bz    {\ensuremath{B^0}\xspace}
\def\B     {\ensuremath{B}\xspace}
\def\Bbar  {\kern 0.18em\overline{\kern -0.18em B}{}\xspace}

\def\Bzb   {\ensuremath{\Bbar^0}\xspace}

\def\BzBzb {\ensuremath{B^0 {\kern -0.16em \Bzb}}\xspace}

\def\jpsi  {\ensuremath{{J\mskip -3mu/\mskip -2mu\psi\mskip 2mu}}\xspace}
\def\psitwos {\ensuremath{\psi{(2S)}}\xspace}

\mathchardef\Upsilon="7107
\def\Y#1S{\ensuremath{\Upsilon{(#1S)}}\xspace}% no space before {...}!

\def\FourS {\Y4S}

\mathchardef\Deltares="7101
\mathchardef\Xi="7104
\mathchardef\Lambda="7103
\mathchardef\Sigma="7106
\mathchardef\Omega="710A
\def\Deltabar   {\kern 0.25em\overline{\kern -0.25em \Deltares}{}\xspace}
\def\Lbar {\kern 0.2em\overline{\kern -0.2em\Lambda\kern 0.05em}\kern-0.05em{}\xspace}
\def\Sigbar{\kern 0.2em\overline{\kern -0.2em \Sigma}{}\xspace}
\def\Xibar{\kern 0.2em\overline{\kern -0.2em \Xi}{}\xspace}
\def\Obar{\kern 0.2em\overline{\kern -0.2em \Omega}{}\xspace}
\def\Nbar{\kern 0.2em\overline{\kern -0.2em N}{}\xspace}
\def\Xb{\kern 0.2em\overline{\kern -0.2em X}{}}

\newcommand{\tev}{\ensuremath{\mathrm{\,Te\kern -0.1em V}}\xspace}
\newcommand{\gev}{\ensuremath{\mathrm{\,Ge\kern -0.1em V}}\xspace}
\newcommand{\mev}{\ensuremath{\mathrm{\,Me\kern -0.1em V}}\xspace}
\newcommand{\kev}{\ensuremath{\mathrm{\,ke\kern -0.1em V}}\xspace}
\newcommand{\ev}{\ensuremath{\mathrm{\,e\kern -0.1em V}}\xspace}
\newcommand{\gevc}{\ensuremath{{\mathrm{\,Ge\kern -0.1em V\!/}c}}\xspace}
\newcommand{\mevc}{\ensuremath{{\mathrm{\,Me\kern -0.1em V\!/}c}}\xspace}
\newcommand{\gevcc}{\ensuremath{{\mathrm{\,Ge\kern -0.1em V\!/}c^2}}\xspace}
\newcommand{\mevcc}{\ensuremath{{\mathrm{\,Me\kern -0.1em V\!/}c^2}}\xspace}
\def\mm   {\ensuremath{\rm \,mm}\xspace}

\def\mum  {\ensuremath{\,\mu\rm m}\xspace}%% mu meter 

\def\invfb   {\ensuremath{\mbox{\,fb}^{-1}}\xspace}
\def\mus  {\ensuremath{\rm \,\mus}\xspace}

\def\ps   {\ensuremath{\rm \,ps}\xspace}

\def\mus        {\ensuremath{\,\mu{\rm s}}\xspace}    %% microsecond
\def\ps         {\ensuremath{{\rm \,ps}}\xspace}  %% picosecond
\def\gsim{{~\raise.15em\hbox{$>$}\kern-.85em
          \lower.35em\hbox{$\sim$}~}\xspace}
\def\lsim{{~\raise.15em\hbox{$<$}\kern-.85em
          \lower.35em\hbox{$\sim$}~}\xspace}

\def\CP                 {\ensuremath{C\!P}\xspace}
\def\to                 {\ensuremath{\rightarrow}\xspace}

\def\pep2{PEP-II}
\def\BF{$B$ Factory}

\def\stwob{\ensuremath{\sin\! 2 \beta   }\xspace}

\def\mistag{\ensuremath{w}\xspace}

\def\deltaz{\ensuremath{{\rm \Delta}z}\xspace}
\def\deltat{\ensuremath{{\rm \Delta}t}\xspace}
\def\deltamd{\ensuremath{{\rm \Delta}m_d}\xspace}

\xspace

\newcommand{\epjc}      [1]  {{Eur.\ Phys.\ Jour.\ C~{\bf #1}}}

\newcommand{\jpl}        [1]  {{Phys.\ Lett.\ {\bf #1}}}      % dbm
\newcommand{\jprl}       [1]  {{Phys.\ Rev.\ Lett.\ {\bf #1}}} % dbm
\newcommand{\pr}        [1]  {{Phys.\ Rev.\ {\bf #1}}}

\def\jetset74   {\mbox{\tt Jetset \hspace{-0.5em}7.\hspace{-0.2em}4}}
\long\def\inst#1{\par\nobreak\kern 4pt\nobreak
    {\it #1}\par\vskip 10pt plus 3pt minus 3pt}

\begin{document}
{\pagestyle{empty}

\begin{flushright}
SLAC-PUB-\SLACPubNumber \\
\babar-PROC-\BABARPubYear/\BABARProcNumber \\
%\babar-PUB-\BABARPubYear/\BABARPubNumber \\
%hep-ex/\LANLNumber \\
November, 2000 \\
\end{flushright}

\par\vskip 4cm

% Title of the paper
\begin{center}
\Large \bf First Measurement of the \CP-Violating Asymmetries with
\Lbabar
\end{center}
\bigskip

\begin{center}
\large 
Yury G. Kolomensky\\
California Institute of Technology \\
Pasadena, California 91125, USA \\
(for the \lbabar\ Collaboration)
\end{center}
\bigskip \bigskip

% Abstract
\begin{center}
\large \bf Abstract
\end{center}
We report on a preliminary  measurement of time-dependent 
\CP-violating asymmetries in  
$\Bz \to \jpsi \KS$ and $\Bz \to \psitwos \KS$  
decays recorded by the \babar\ detector at the \pep2\
asymmetric \BF\ at SLAC.  
The data sample consisted of 9.0\invfb\  collected 
at the \FourS\ resonance and 0.8\invfb\ off-resonance.  
One of the pair of neutral \B\ mesons produced at the \FourS was
fully reconstructed, while 
the flavor of the other neutral \B\
meson was tagged at the time of its decay.  
The value of the asymmetry amplitude, \stwob, was determined from
a maximum likelihood fit 
to the time distribution of 120 tagged candidates to be 
$\stwob=0.12\pm 0.37 {\rm (stat.)} \pm 0.09 {\rm (syst.)}$
(preliminary).

\vfill
\begin{center}
Contributed to the Proceedings of the DPF2000 Meeting \\
of the Division of Particles and Fields of the American Physics Society,\\
8/9/2000---8/12/2000, Columbus, OH
\end{center}

\vspace{1.0cm}
\begin{center}
{\em Stanford Linear Accelerator Center, Stanford University, 
Stanford, CA 94309} \\ \vspace{0.1cm}\hrule\vspace{0.1cm}
Work supported in part by Department of Energy contract DE-AC03-76SF00515.
\end{center}
}
\newpage

The \CP-violating phase of the three-generation
Cabibbo-Kobayashi-Maskawa ({\mbox{CKM}) 
quark mixing matrix can provide an
elegant explanation of the \CP-violating effects seen in decays of
neutral $K$ mesons\cite{EpsilonK}.  The unitarity relations between
the elements of the CKM matrix
can be expressed as six triangles of equal area in the complex plane.
A nonzero area\cite{Jarlskog} directly implies the existence of a
\CP-violating CKM phase.  The most experimentally accessible of the
unitarity relations involves elements $V_{ub}$ and $V_{td}$, 
and is known as the Unitarity
Triangle, where angles, and hence size of \CP-violating asymmetries
are expected to be large\cite{PhysBook}. Observing the \CP-violating
asymmetry can thus provide a crucial test of the Standard Model.  

The \CP-violating asymmetry in  $\b \to \ccbar \s$ decays of the \Bz\ meson 
such as  $\Bz/\Bzb \to \jpsi \KS$ 
(or $\Bz/\Bzb \to \psitwos \KS$) 
is caused by the interference between mixed and unmixed decay amplitudes.
With little theoretical uncertainty, the phase difference 
between these amplitudes is equal to twice the angle 
$\beta$ of the
Unitarity Triangle.  
In \epem\ storage rings operating at  the \FourS\ resonance a \BzBzb\ pair
produced in \FourS\ decay 
evolves in a coherent $P$-wave until one of the \B\ mesons decays. 
For this measurement, one of the \B\ mesons ($B_{CP}$) 
is fully reconstructed in a \CP\ 
eigenstate ($\jpsi \KS$ or $\psitwos \KS$).
If the other \B\ meson ($B_{tag}$) is determined
to decay to a state of known flavor at a certain time $t_{tag}$, 
$B_{CP}$  is {\it at that time} known to be of the opposite flavor.
By measuring the proper time interval $\deltat = t_{CP} - t_{tag}$ from the $B_{tag}$ decay time 
to the decay of the $B_{CP}$, it is possible to determine the time evolution 
of the initially pure \Bz\ or \Bzb\ state.  
The experimental 
time-dependent decay rate into the $B_{CP}$ final state is given by 
\begin{equation}
\label{eq:TimeDep}
 {\cal F}_\pm = {\frac{1}{4}}\, \Gamma \, {\rm e}^{ - \Gamma \left| \deltat \right| }
\, \left[  \, 1 \, \pm \, {\cal {D}} \sin{ 2 \beta } \times \sin{ \deltamd \, \deltat } \,  \right]\ \otimes 
{\cal {R}}( \, \deltat \, ; \, \hat {a} \, ) \ ,
\end{equation}
where the $+$ or $-$  sign 
indicates whether the 
$B_{tag}$ is tagged as a \Bz\ or a \Bzb, respectively.  
The ``dilution factor'' ${\cal {D}}$ is given by 
$ {\cal {D} } = 1 - 2 \mistag$, 
where $\mistag$ is the mistag fraction, {\it i.e.}, the 
probability that the 
flavor of the tagging \B\ is identified incorrectly. The 
term ${\cal {R}}$ accounts for the finite detector resolution, 
where $\hat {a}$ represents the set of parameters that describe the 
resolution function.  

\section{Sample Selection}

For this analysis we used a sample of $9.8 \invfb$ of data recorded
by the \babar\ detector~\cite{BabarPub0018} 
between January 2000 and the beginning of July 2000, 
of which $0.8 \invfb$ was recorded 40\mev\ below the \FourS\ 
resonance (off-resonance data). At \pep2, $B$ mesons are produced in
the asymmetric collisions of $9$ GeV electrons and $3.1$ GeV
positrons, and have an average boost along $z$ direction of $\langle
\beta\gamma\rangle = 0.56$.

$\jpsi$ candidates were identified through their decays into $\epem$
and $\mu^+ \mu^-$, and $\psitwos$ candidates were reconstructed in
$\epem$, $\mu^+ \mu^-$, and $\jpsi \pi^+ \pi^-$
modes\cite{BabarPub0005}. In the electron modes of $\jpsi$ and $\psitwos$,
at least one electron was required to be positively identified. An
algorithm for the recovery of Bremsstrahlung
photons\cite{BabarPub0005} was used if both
electrons were positively identified. Similarly, for the $\mu^+ \mu^-$
candidates, at least one muon was required to pass particle
identification criteria. The charmonium states then were selected in a
window around their expected mass\cite{BabarPub0005}, and were combined
with \KS\ candidates reconstructed through their decays into 
 $\pi^+ \pi^-$ and  $\pi^0 \pi^0$ states\cite{BabarPub0005} to form
$B_{CP}$ candidates.  
$B_{CP}$ candidates were identified with a pair of nearly uncorrelated 
kinematic variables: the 
difference ${\rm \Delta}E$ between the energy of the $B_{CP}$ candidate and the 
beam energy in the center-of-mass frame, 
and the beam-energy substituted mass $m_{\rm SE}$~\cite{BabarPub0018}.
The signal region was defined by 
$ 5.270 \gevcc < m_{\rm SE} <  5.290 \gevcc$ and
an approximately three-standard-deviation cut on ${\rm \Delta}E$
(typically $\left| {\rm \Delta} E \right| < 35\mev$).
The \CP\ sample used  in this 
analysis was composed of 168 candidates:  
121 in the $\jpsi \KS$ ($\KS \to \pi^+ \pi^-$) channel, 
19 in the  $\jpsi \KS$ ($\KS \to \pi^0 \pi^0$) channel and 
28 in the $\psitwos \KS$ ($\KS \to \pi^+ \pi^-$) channel. 

\section{Time Resolution Function}

All tracks in the event, excluding those from $B_{CP}$, were fit to a
common vertex using an iterative procedure, which at each step removes
a track with the worst $\chi^2$ greater than 6. Information on the
direction of the $B_{CP}$, derived from the kinematics of the
reconstructed $B_{CP}$ and the average position of the interaction
point, was also used in the fit. In order  to reduce bias,
all pairs of tracks that can be
reconstructed as long-lived neutral particles ($V^0$s), were replaced
by the corresponding $V^0$ candidates in the fit. 
Events were rejected if either $B_{CP}$ or $B_{tag}$ does not converge,
or if reconstructed $\deltaz$ or its uncertainty were 
large($\left| \deltaz \right| > 3\mm$ or $\sigma_{\deltaz}>400\mum$).

The time resolution function 
is described accurately by the sum of
 three Gaussian distributions, which has six independent parameters:
\begin{eqnarray}
{\cal {R}}( \, \deltat ; \, \hat {a} \,  ) &=&  \sum_{i=1}^{3} { \, \frac{f_i}{\sigma_i\sqrt{2\pi}} \, {\rm exp} \left(  - ( \deltat-\delta_i )^2/2{\sigma_i }^2   \right) } \, \, .
\end{eqnarray}
\par
A fit to the time resolution function in
Monte Carlo simulated events indicated 
that most of the events ($f_1 \approx 1-f_2 = 70\%$) were in the core 
Gaussian, which has a width $\sigma_1 \approx 0.6 \ps$.
The wide Gaussian had a width  $\sigma_2 \approx 1.8\ps$. 
Tracks from forward-going charm decays included in the reconstruction
of the $B_{tag}$ vertex 
introduced a small bias, 
$\delta_1 \approx -0.2 \ps$, for the core Gaussian.  The third
Gaussian, used to parameterize events at very large values of
$\deltaz$ due to poorly reconstructed vertices, had a fixed width of 8
ps and a fraction $f_w\sim 1\%$\cite{BabarPub0001}. 
In likelihood fits, the widths of the
first and second Gaussian were parameterized by 
event-by-event error $\sigma_{\deltat}$ from the vertex fits and two
scale factors 
${\cal S}_1$ and ${\cal S}_2$
($\sigma_1={\cal S}_1 \times \sigma_{\deltat}$ and 
$\sigma_2={\cal S}_2 \times \sigma_{\deltat}$). Parameters ${\cal S}_2=2.1$
and $f_2=0.25$ were determined from Monte Carlo, bias of the second
Gaussian $\delta_2$ was fixed at 0 ps, and three remaining parameters 
$\hat {a} = \{ {\cal S}_1, \delta_1, f_{w} \}$ were determined from the
observed vertex distributions in the hadronic $B^0$ decays. 
The time resolution is dominated by the precision 
of the $B_{tag}$ vertex position, and we found  no significant 
differences in the Monte Carlo simulation of 
the resolution function parameters 
for the various fully reconstructed decay modes\cite{BabarPub0001}. 

\section{\boldmath $B$ flavor tagging}

Each event with a \CP\ candidate was assigned a $\Bz$ or $\Bzb$ tag if 
the rest of the event ({\it i.e.,} with the daughter
tracks of the $B_{CP}$ removed) satisfies the criteria for one of several 
tagging categories. 
Three tagging categories
relied on the presence of a fast lepton or 
one or more charged kaons in the event.  
Two categories, called neural network categories, were based upon
the output value of a neural network algorithm applied to events 
that have not already been assigned to lepton or kaon  
tagging categories.

The mistag fractions $\mistag_i$ were measured directly
for each tagging category by studying the probability of flavor 
mixing in events 
in which one \Bz\ candidate, called the $B_{rec}$, was fully reconstructed 
in a flavor eigenstate mode\cite{BabarPub0008}. 
Both integrated and time-dependent analyses of the 
fraction of mixed events are sensitive to the to the mistag
fraction. In the time-dependent analysis, the probability density
functions of unmixed $(+)$ and mixed $(-)$ events were defined similar
to Eq.~(\ref{eq:TimeDep})
\begin{equation}
 {\cal H}_\pm =
{\frac{1}{4}}\, \Gamma \, {\rm e}^{ - \Gamma \left| \deltat \right| }
\, \left[  \, 1 \, \pm \, {\cal {D}} \times \cos{ \deltamd \, \deltat } \,  \right]
 \otimes 
{\cal {R}}( \, \deltat; \, \hat {a} )  \, .
\end{equation}
These functions were used in an maximum maximum likelihood fit to the
large samples of fully reconstructed $B^0$ decays into hadronic and
semileptonic modes to determine mistag fractions 
$\mistag_i = \frac{1}{2}(1-{\cal D}_i)$
for each tagging category $i$. We find the tagging efficiency 
$\varepsilon = (76.7\pm0.5)\%$ and the effective tagging efficiency 
$Q = \varepsilon \, \left( 1 - 2\mistag \right)^2 = (27.9 \pm 0.5) \%$
(statistical errors only)\cite{BabarPub0008}. Out of 168 \CP\
candidates, 120 were tagged: 70 as \Bz\ and 50 as \Bzb. 

\section{Extracting \boldmath \stwob}

The value of \stwob\ was
determined from an unbinned maximum likelihood fit to the
time-dependent distribution of tagged $B_{CP}$ decays\cite{BabarPub0001}.
The fitting procedure was extensively tested in ``toy'' Monte Carlo
simulations, where effects backgrounds, different resolution function
parameterizations, and behavior of the fit on small statistical
samples were studied. We also verified the reconstruction and fitting
procedure on a large sample of events produced with the full \babar\
GEANT3 detector simulation. 

The analysis was also validated on the various data control samples
where the \CP\ asymmetry was expected to be zero. An ``apparent \CP\
asymmetry'' was studied on a sample of  $B^+ \to \jpsi K^+$ events
and events with self-tagged  $\Bz\to \jpsi K^{*0}$  ($K^{*0} \to K^+ \pi^-$)
decays.  We also use the event samples with fully-reconstructed
candidates in charged or neutral hadronic modes. In all cases, the
apparent \CP\ asymmetry was found to be consistent with
zero\cite{BabarPub0001}.

We have adopted a blind analysis for the extraction of \stwob\ 
in order to eliminate possible experimenter's bias.  
We used a technique 
that hides not only the result of the unbinned 
maximum likelihood fit, but also the visual \CP\ asymmetry
in the \deltat\ distribution.  
The error on the asymmetry was not hidden.
With these techniques, numerous systematic studies could be performed
while keeping the numerical value of \stwob\  hidden.  
The analysis procedure for extracting 
\stwob\ was frozen, and the
data sample fixed, prior to unblinding. 

Using the maximum-likelihood fit to the full data sample of $\Bz/\Bzb \to \jpsi \KS$ and $\Bz/\Bzb \to \psitwos \KS$ 
events, we determined\cite{BabarPub0001}
\begin{equation}
\stwob=0.12\pm 0.37 {\rm (stat.)} \pm 0.09 {\rm (syst.)}\, {\rm(preliminary)} .
\end{equation}
Systematic errors arise from uncertainties in 
input parameters to the maximum likelihood fit, 
incomplete knowledge of the time resolution function, 
uncertainties in the mistag fractions, 
and possible limitations in the analysis procedure.  

\section{Conclusions and prospects}

We have presented the  first measurement of the \CP-violating 
asymmetry parameter \stwob\ in the $B$ meson system by \babar
Collaboration. Our measurement is consistent with the presently
available data\cite{worldSin2b}, and also agrees within two standard
deviations with the value determined by other constraints of the
Unitarity Triangle\cite{PhysBook}. While the current experimental
uncertainty on \stwob\ is large, the next few years will bring significant
improvements in precision. Measurements of \CP-violating asymmetries
in other final states are also underway, and will help constrain the
Standard Model description of \CP\ violation.

%\nonumsection{References}


\begin{thebibliography}{000}
\bibitem{EpsilonK}
J.\ H.\ Christenson {\em et al.}, \jprl {13}, 138 (1964); \\
NA31 Collaboration, G.\ D.\ Barr {\em et al.}, \jpl {317}, 233 (1993);\\
E731 Collaboration, L.\ K.\ Gibbons {\em et al.}, \jprl {70}, 1203 (1993).

\bibitem{Jarlskog}
C.\ Jarlskog, in {\em CP Violation}, ed. C. Jarlskog, World
Scientific, Singapore (1988).

\bibitem{PhysBook}
P.\ H.\ Harrison and H.\ R.\ Quinn, eds., {\em The \babar\ physics book}, 
Report SLAC-R-504 (1998).

\bibitem{BabarPub0008}
\babar\ Collaboration, B.\ Aubert {\em et al.},
Preprint \babar-CONF-00/08, SLAC-PUB-8530, hep-ex/0008052,
submitted to the XXX$^{th}$
International Conference on High Energy Physics, Osaka, Japan;
S.\ Rahatlou, in these proceedings. 

\bibitem{BabarPub0001}
\babar\ Collaboration, B.\ Aubert {\em et al.},
Preprint \babar-CONF-00/01, SLAC-PUB-8540, hep-ex/0008048, 
submitted to the XXX$^{th}$
International Conference on High Energy Physics, Osaka, Japan.

\bibitem{BabarPub0018}
\babar\ Collaboration, B.\ Aubert {\em et al.},
Preprint \babar-CONF-00/17, SLAC-PUB-8539, 
submitted to the XXX$^{th}$
International Conference on High Energy Physics, Osaka, Japan.

\bibitem{BabarPub0005}
\babar\ Collaboration, B.\ Aubert {\em et al.},
\babar-CONF-00/05, SLAC-PUB-8527, hep-ex/0008051,
submitted to the XXX$^{th}$
International Conference on High Energy Physics, Osaka, Japan;
G.\ Dubois-Felsmann, in these proceedings. 

\bibitem{worldSin2b}
OPAL Collaboration, K.\ Ackerstaff {\em et al.}, \epjc{5}, 379 (1998);\\
CDF Collaboration, T.\ Affolder {\em et al.}, \pr{D61}, 072005 (2000);
W.\ Trischuk, in these proceedings;\\
ALEPH Collaboration, ALEPH 99-099 CONF 99-54 (1999);
B.\ Petersen, in these proceedings;\\
Belle Collaboration, H.\ Aihara {\em et al.},
Preprint hep-ex/0010008,
submitted to the XXX$^{th}$
International Conference on High Energy Physics, Osaka, Japan;
J.\ Rodriguez, in these proceedings.

\end{thebibliography}
\end{document}